\documentclass[12pt,a4paper,dvips,final]{article}
\usepackage{epsfig,wrapfig,times,mathptm} 
\newcommand{\degr}{^\circ}
\renewcommand{\section}[1]{\vspace{6pt} \noindent\mbox{#1} \newline \noindent}
\renewcommand{\subsection}[1]{\vspace{6pt} \noindent\mbox{\underline{#1}} 
\newline \noindent}
\renewcommand{\subsubsection}[1]{\vspace{6pt} \noindent\mbox{\underline{#1}}
\noindent}

\newfont{\sansb}{cmssbx10}
\newfont{\sans}{cmss10}

\def\Dunits{$\times10^{28}$ cm$^2$s$^{-1}$}
\begin{document}
{\small Proc. 25th Int. Cosmic Ray Conference, Durban, 1997, v.4,
   p.257--260 \vspace{-24pt}\\}     

{\center \LARGE NUMERICAL MODELS FOR COSMIC RAY \\PROPAGATION AND
   GAMMA RAY PRODUCTION \vspace{6pt}\\}
A.W.\ Strong$^*$, I.V.\ Moskalenko$^{*\dagger}$, and V.\ Sch\"onfelder$^*$ 
\vspace{6pt}\\
{\it $^*$Max-Planck-Institut f\"ur extraterrestrische Physik,
              Postfach 1603, D-85740 Garching, Germany\\
$^{\dagger}$Institute for Nuclear Physics,
             M.V.\ Lomonosov Moscow State University,
             119 899 Moscow, Russia \vspace{-12pt}\\}
{\center ABSTRACT\\}
An extensive program for the calculation of galactic cosmic-ray
propagation has been developed.  Primary and secondary nucleons, primary
and secondary electrons, and secondary positrons are included. The basic
spatial propagation mechanisms are (momentum-dependent) diffusion,
convection, while in momentum space energy loss and diffusive
reacceleration are treated.  Fragmentation and energy losses are
computed using realistic  distributions for the interstellar gas and
radiation fields.

\setlength{\parindent}{1cm}
\section{INTRODUCTION}
The main motivation for developing this code (Strong and Moskalenko
1997) is the prediction of diffuse Galactic gamma rays for comparison
with data from the CGRO instruments EGRET, COMPTEL and OSSE. This is a
development of the work described in Strong and Youssefi (1995).  More
generally the idea is to develop a model which self-consistently
reproduces observational data of many kinds related to cosmic-ray origin
and propagation: direct measurements of nuclei, electrons and positrons,
gamma rays, and synchrotron radiation. These data provide many
independent constraints on any model and our approach is able to take
advantage of this since it must be consistent with all types of
observation.  We emphasize also the use of realistic astrophysical input
(e.g.\ for the gas distribution) as well as theoretical developments
(e.g.\ reacceleration).  The code is sufficiently general that new
physical effects can be introduced as required.  The basic procedure is
first to obtain a set of propagation parameters which reproduce the
cosmic ray $B/C$ ratio, and the spectrum of secondary positrons;  the
same propagation conditions are then applied to primary electrons.
Gamma-ray and synchrotron emission are then evaluated. Models both with
and without reacceleration are considered.

\section{DESCRIPTION OF THE MODEL}
The models are three dimensional with cylindrical symmetry in the
Galaxy, the basic coordinates being $(R,z,p)$ where $R$ is
Galactocentric radius, $z$ is the distance from the Galactic plane and
$p$ is the total particle momentum.  The numerical solution of the
transport equation is based on a Crank-Nicholson implicit second-order
scheme. In the models the  propagation region is bounded by $z=z_h$
beyond which free escape is assumed.  A value $z_h=3$ kpc  has been
adopted since this is within the  range which is consistent with studies
of $^{10}Be/Be$ and synchrotron radiation. For a given $z_h$ the
diffusion coefficient as a function of momentum is determined by $B/C$
for the case of no reacceleration; with reacceleration on the other hand
it is the reacceleration strength (related to the Alfven speed v$_A$)
which is determined by $B/C$.  Reacceleration provides a natural
mechanism to reproduce the $B/C$ ratio without an ad-hoc form for the
diffusion coefficient (e.g., Heinbach and Simon 1995, Seo and Ptuskin
1994).  The spatial diffusion coefficient for the case {\it without}
reacceleration is $D=\beta D_0$ below rigidity $\rho_0$, $\beta
D_0(\rho/\rho_0)^\delta$ above rigidity $\rho_0$, where $\beta$ is the
particle speed.  The spatial diffusion coefficent {\it with}
reacceleration assumes a Kolmogorov spectrum of weak MHD turbulence so
$D=\beta D_0(\rho/\rho_0)^\delta$ with $\delta=1/3$ for all rigidities.
For this case  the momentum-space diffusion coefficient is related to
the spatial coefficient (Seo and Ptuskin 1994).  The injection spectrum
of nucleons is assumed to be a power law in momentum for  the different
species, $dq(p)/dp \propto p^{-\gamma}$ for the injected {\it density},
corresponding to an injected {\it flux} $dF(E_k)/dE_k \propto
p^{-\gamma}$ where $E_k$ is the kinetic energy per nucleon.

The entire calculation is performed with momentum as the kinematic
variable.  The interstellar hydrogen distribution uses {\it HI} and {\it
CO} surveys and information on the ionized component; the Helium
fraction of the gas is taken as 0.11 by number.  The interstellar
radiation field for inverse Compton losses is based on stellar
population models and IRAS and COBE data, plus the cosmic microwave
background. The magnetic field is assumed to have the form
$6\,e^{-(|z|/5{\rm kpc}) - (R/20{\rm kpc})}$ $\mu$G.  Energy losses for
electrons by ionization, Coulomb, bremsstrahlung, inverse Compton and
synchrotron are included, and for nucleons by ionization and Coulomb
interactions.  The distribution of cosmic-ray sources is chosen to
reproduce the cosmic-ray distribution determined by analysis of EGRET
gamma-ray data (Strong and Mattox 1996).  The bremsstrahlung and inverse
Compton gamma rays are computed self-consistently from the gas and
radiation fields used for the propagation. The $\pi^0$-decay gamma rays
are calculated directly from the proton and Helium spectra using the
method of Dermer (1986). The secondary nucleon and secondary $e^\pm$
source functions are computed from the propagated primary distribution
and the gas distribution, and the anisotropic distribution of $e^\pm$ in
the $\mu^\pm$ system was taken into account.

\def\figheight{60mm}
\def\figwidth{80mm}
\begin{wrapfigure}[9]{r}[1mm]{80mm}
   \begin{picture}(80,60)(0,0)
   \put(-5,13){\makebox(80,60)[l]%
{\psfig{file=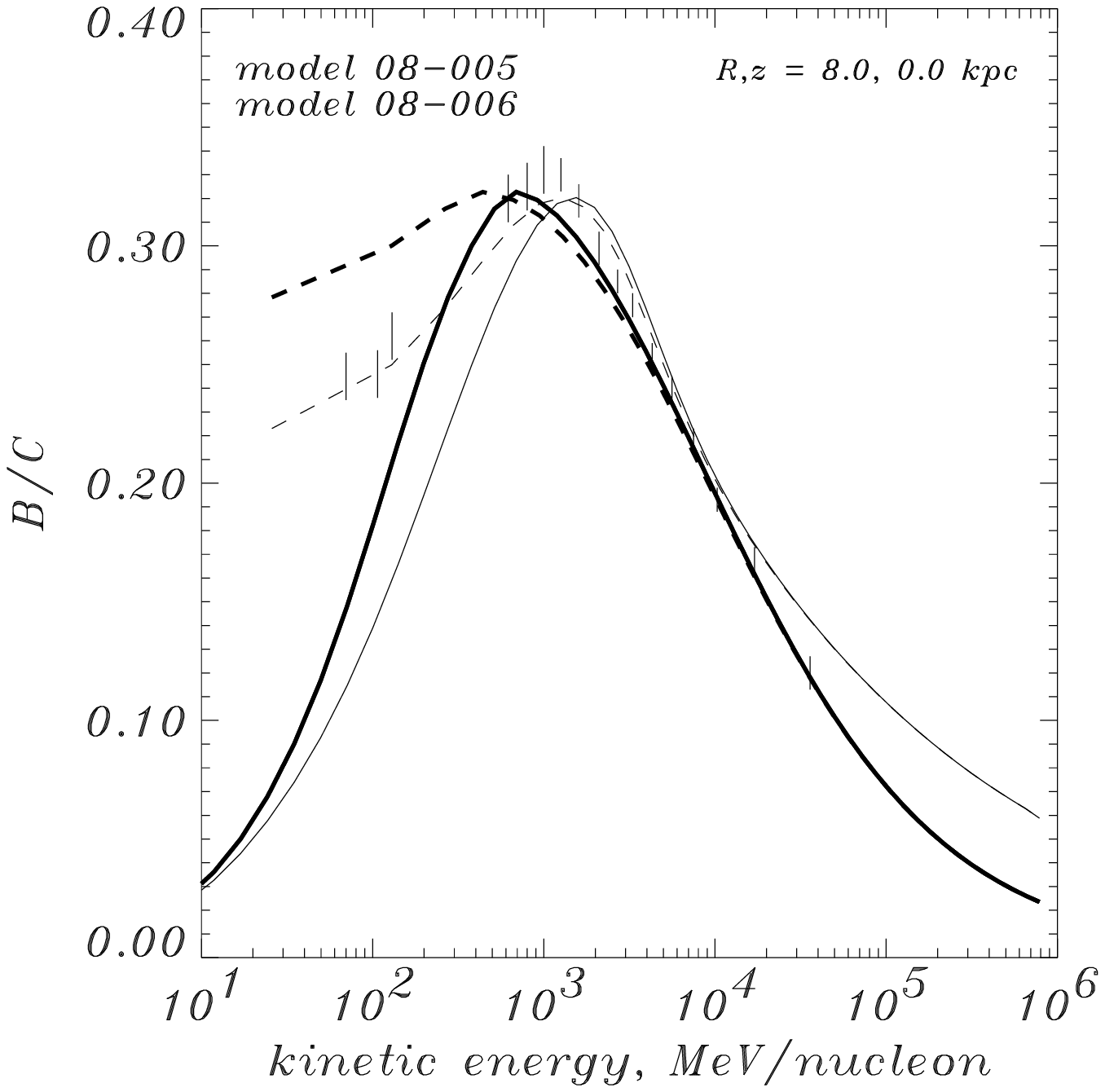,height=\figheight,width=\figwidth,clip=}}}
   \end{picture}
\end{wrapfigure}

\section{ILLUSTRATIVE RESULTS}
Some  results obtained are shown in the Figures.
\medskip

\noindent {\it Fig.~1: Secondary nucleons.}
The energy dependence of the $B/C$ ratio can be reproduced with $D_0 =
2.0$\Dunits, $\delta=0.6$, $\rho_0= 3$ GV/c without reacceleration
(thick line) and $D_0 = 4.2$\Dunits, v$_A=20$ km s$^{-1}$ with
reacceleration (thin line).  Dashed lines are modulated to 500 MV. Data
from Webber et al.\ (1996).

\def\figheight{65mm}
\def\figwidth{80mm}
\begin{figure}[h]
   \begin{picture}(170,65)(0,0)
      \put(0,0){\makebox(80,65)[l]%
{\psfig{file=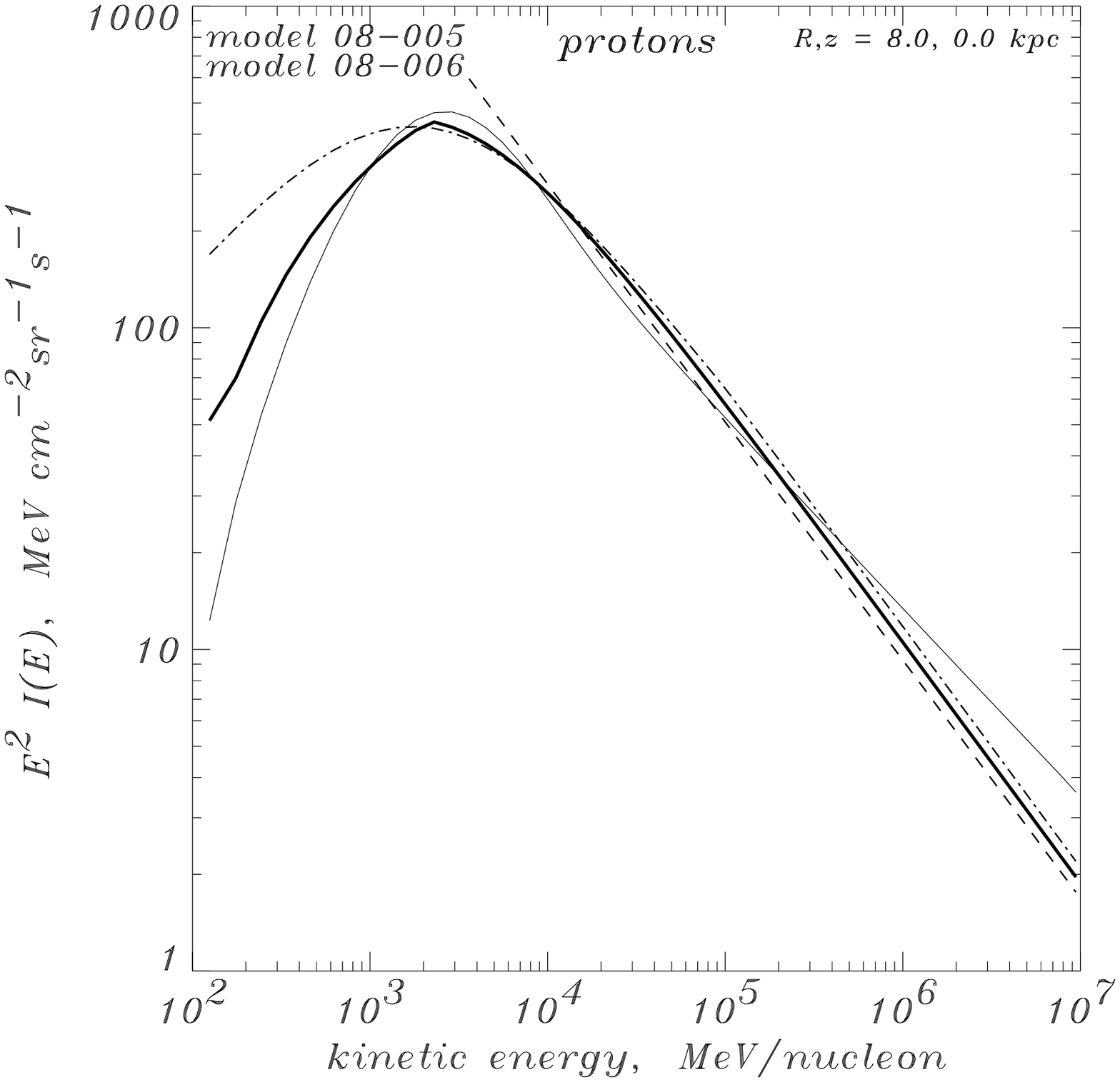,height=\figheight,width=\figwidth,clip=}}}
      \put(85,0){\makebox(80,65)[l]%
{\psfig{file=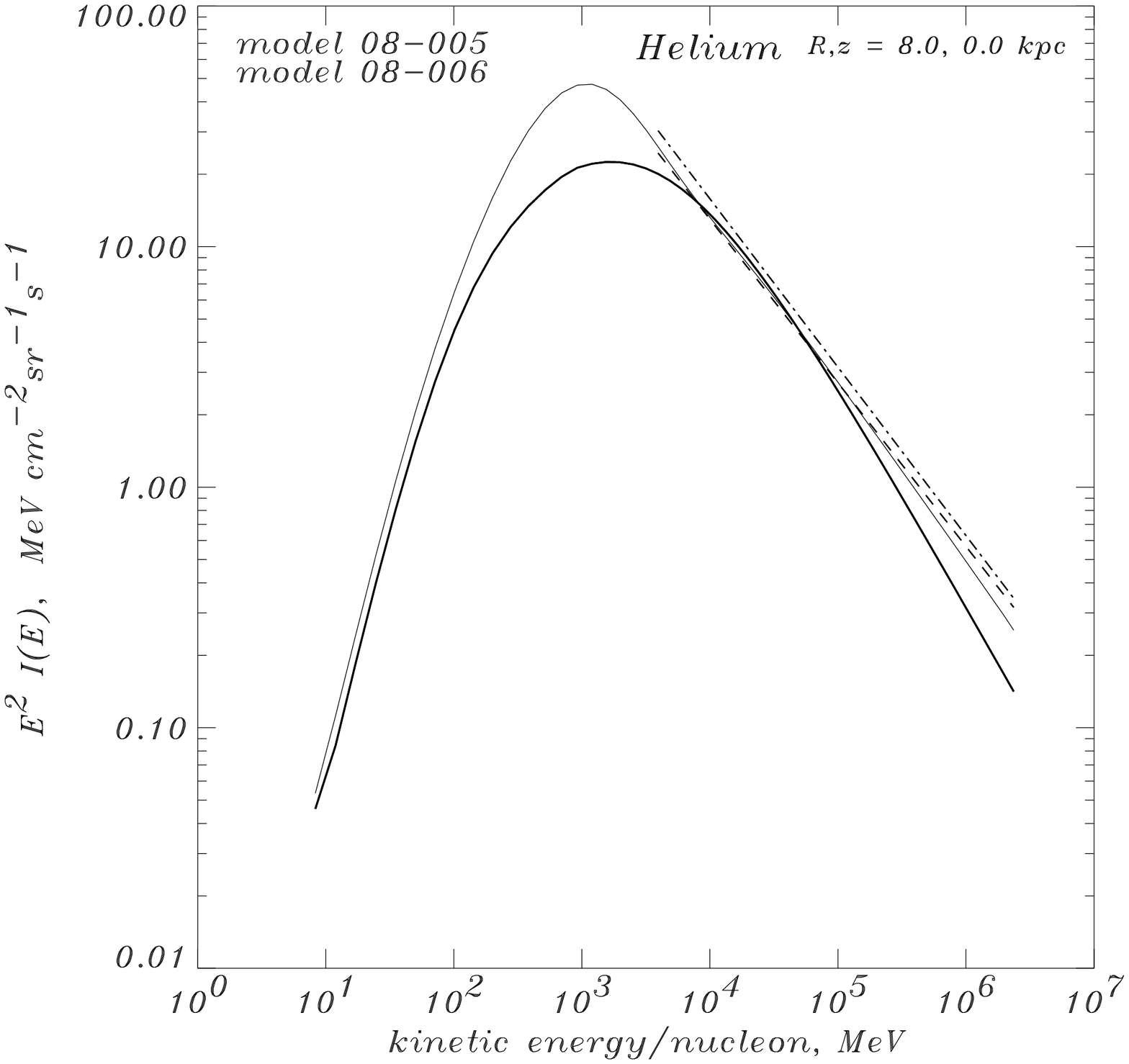,height=\figheight,width=\figwidth,clip=}}}
   \end{picture}
{\it Fig.~2:  Left panel:}
the local proton spectrum for injection index 2.15 (thin solid line),
2.25 (thick solid line) without and with reacceleration respectively,
compared with the measured `interstellar' spectrum (dashed line: Seo et
al.\ 1991, dashed-dot line: Mori 1997).
{\it Right panel:}
the Helium spectrum with injection index 2.25  (thin
solid line), 2.45 (thick solid line)  without and with reacceleration
respectively, compared with the measured `interstellar' spectrum (dashed
line: Seo et al.\ 1991, dashed-dot line: Engelmann et al.\ 1990).  The
spectra are well reproduced up to 100 GeV.
\vskip -2mm
\end{figure}

{\it Secondary positrons and electrons:}
Using the primary proton and Helium spectra, the propagation of
secondary $e^\pm$ has been computed. The $e^+$ spectrum (Fig~3) and the
positron fraction (Fig~4) agrees well with the most recent data
compilation for 0.1--10 GeV; for more details see Moskalenko and Strong
(1997).

{\it Primary electrons:}
The spectrum of primary electrons is also shown in Fig~3.  The adopted
electron injection spectrum has a power law index --2.1 up to 10 GeV;
this is chosen using the constraints from synchrotron and from gamma
rays.  The spectrum is consistent with the direct measurements around 10
GeV where solar modulation is small and it also satisfies the
constraints from ${e^+}\over {e^++e^-}$.  Above 10 GeV a break is
required to at least --2.4 for agreement with direct measurements (which
may however not be necessary if local sources dominate the  directly
measured high-energy electron spectrum).

\def\figheight{50mm}
\def\figwidth{80mm}
\def\height{70mm}
\begin{figure}[thb]
   \begin{picture}(170,5)(0,0)
      \put(5,3){\makebox(20,0)[l]%
         {\small $E^2 I$, MeV cm$^{-2}$ sr$^{-1}$ s$^{-1}$}}
      \put(90,3){\makebox(20,0)[l]%
         {\small $E^2 I$, MeV cm$^{-2}$ sr$^{-1}$ s$^{-1}$}}
    \end{picture}

   \begin{picture}(170,70)(0,0)
\put(0,0){\makebox(80,70)[l]{\psfig{file=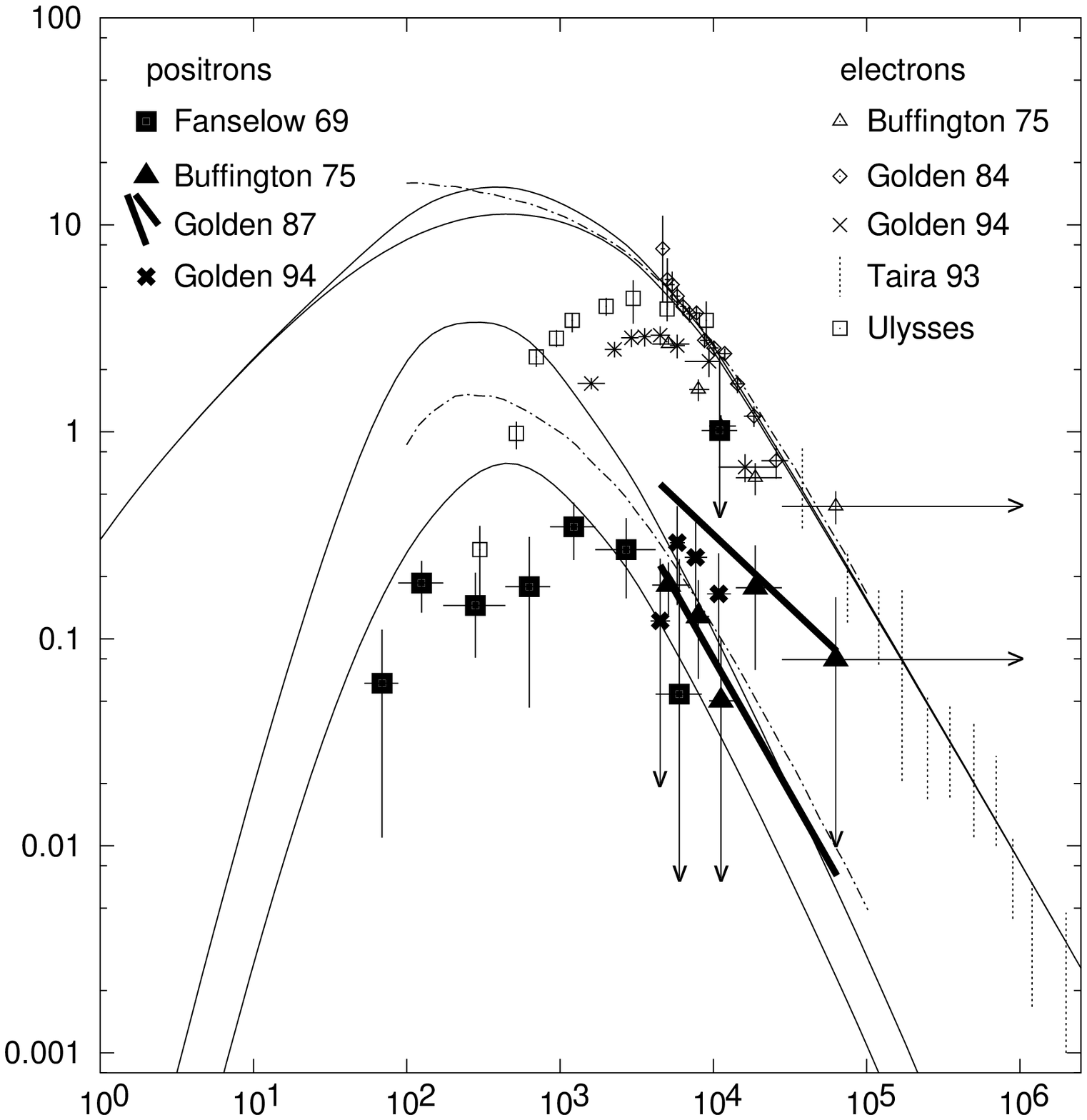,%
   height=\height,width=\figwidth,clip=}}}

\put(85,0){\makebox(80,70)[l]{\psfig{file=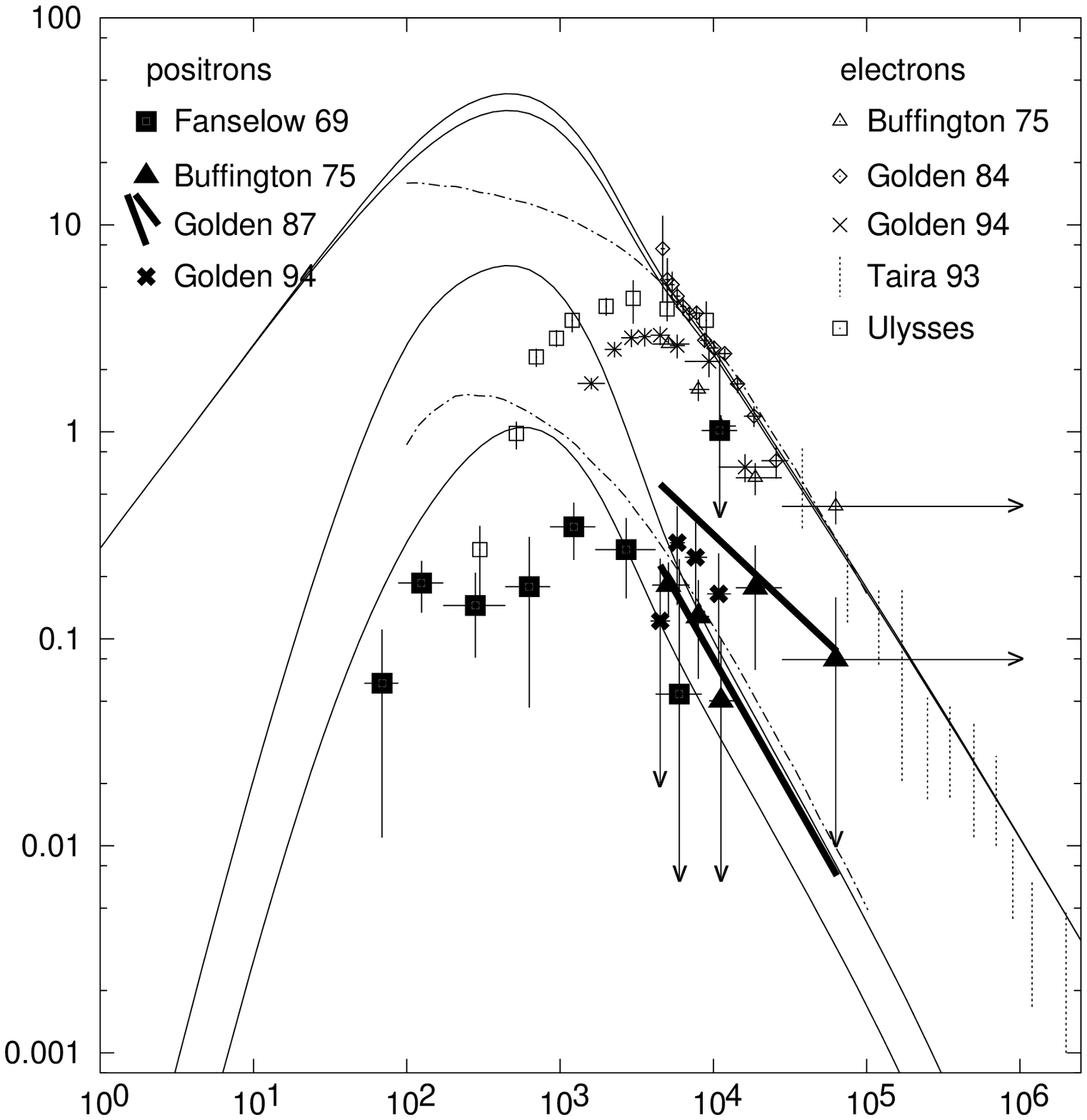,%
   height=\height,width=\figwidth,clip=}}}
      \put(17,30){\makebox(5,2)[c]{\scriptsize sec. $e^+$}}
      \put(25,10){\makebox(5,2)[c]{\scriptsize sec. $e^-$}}
      \put(43,58){\makebox(5,2)[c]{\scriptsize Total}}
      \put(102,30){\makebox(5,2)[c]{\scriptsize sec. $e^+$}}
      \put(110,10){\makebox(5,2)[c]{\scriptsize sec. $e^-$}}
      \put(128,63){\makebox(5,2)[c]{\scriptsize Total}}
   \end{picture}

   \begin{picture}(170,6)(0,-1)
      \put(25,0){\makebox(40,3)[c]{\small Energy, MeV}}
      \put(110,0){\makebox(40,3)[c]{\small Energy, MeV}}
    \end{picture}
{\it Fig.~3:  Spectra of secondary positrons and electrons, and of
primary electrons. Full lines: our model with no reacceleration (left
panel) and with reacceleration (right panel). Dash-dotted lines
(Protheroe 1982): lower is his leaky-box prediction for $e^+$, upper is
his electron spectrum.}

\def\height{130mm}
   \begin{picture}(170,5)(0,0)
      \put(5,0){\makebox(20,0)[l]{\small $e^+/(e^+ +e^-)$}}
      \put(90,0){\makebox(20,0)[l]{\small $e^+/(e^+ +e^-)$}}
   \end{picture}

   \begin{picture}(170,80)(5,8)
\put(0,0){\makebox(80,70)[l]{\psfig{file=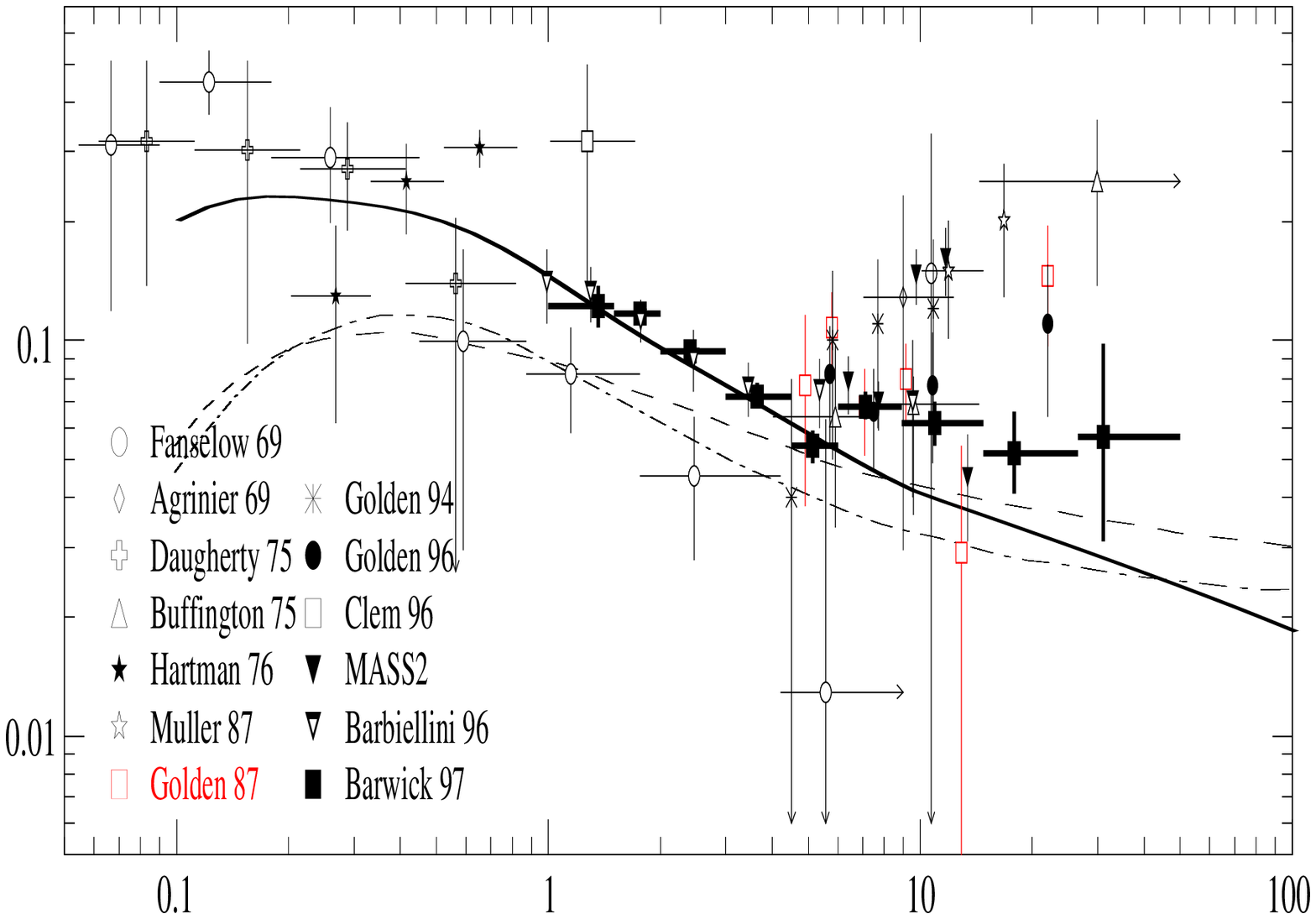,height=\height,clip=}}}

\put(85,0){\makebox(80,70)[l]{\psfig{file=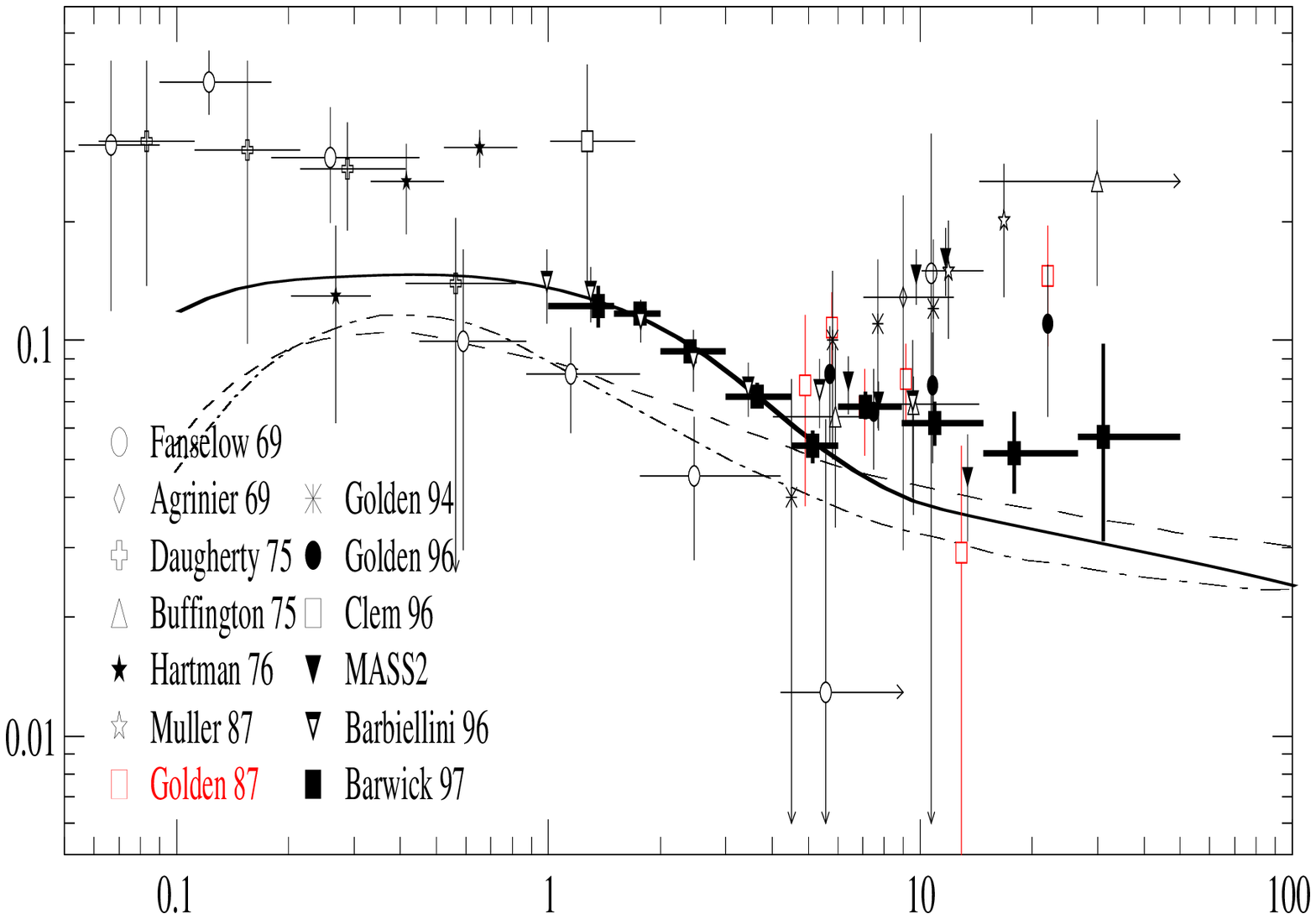,height=\height,clip=}}}
   \end{picture}

   \vskip -21mm
   \begin{picture}(170,6)(0,-1)
      \put(25,0){\makebox(40,3)[c]{\small Energy, GeV}}
      \put(110,0){\makebox(40,3)[c]{\small Energy, GeV}}
   \end{picture}
{\it Fig.~4: Positron fraction for model with no reacceleration (left
panel) and with reacceleration (right panel). Dashed and dash-dotted
lines: Protheroe (1982) predictions of the leaky-box and diffusive halo
models respectively. The data collection is taken from Barwick et al.\
(1997).}
\end{figure}
\vskip -3mm

\def\figheight{65mm}
\def\figwidth{85mm}
\begin{figure}[thb]
   \begin{picture}(170,65)(0,0)
      \put(0,0){\makebox(80,65)[l]%
{\psfig{file=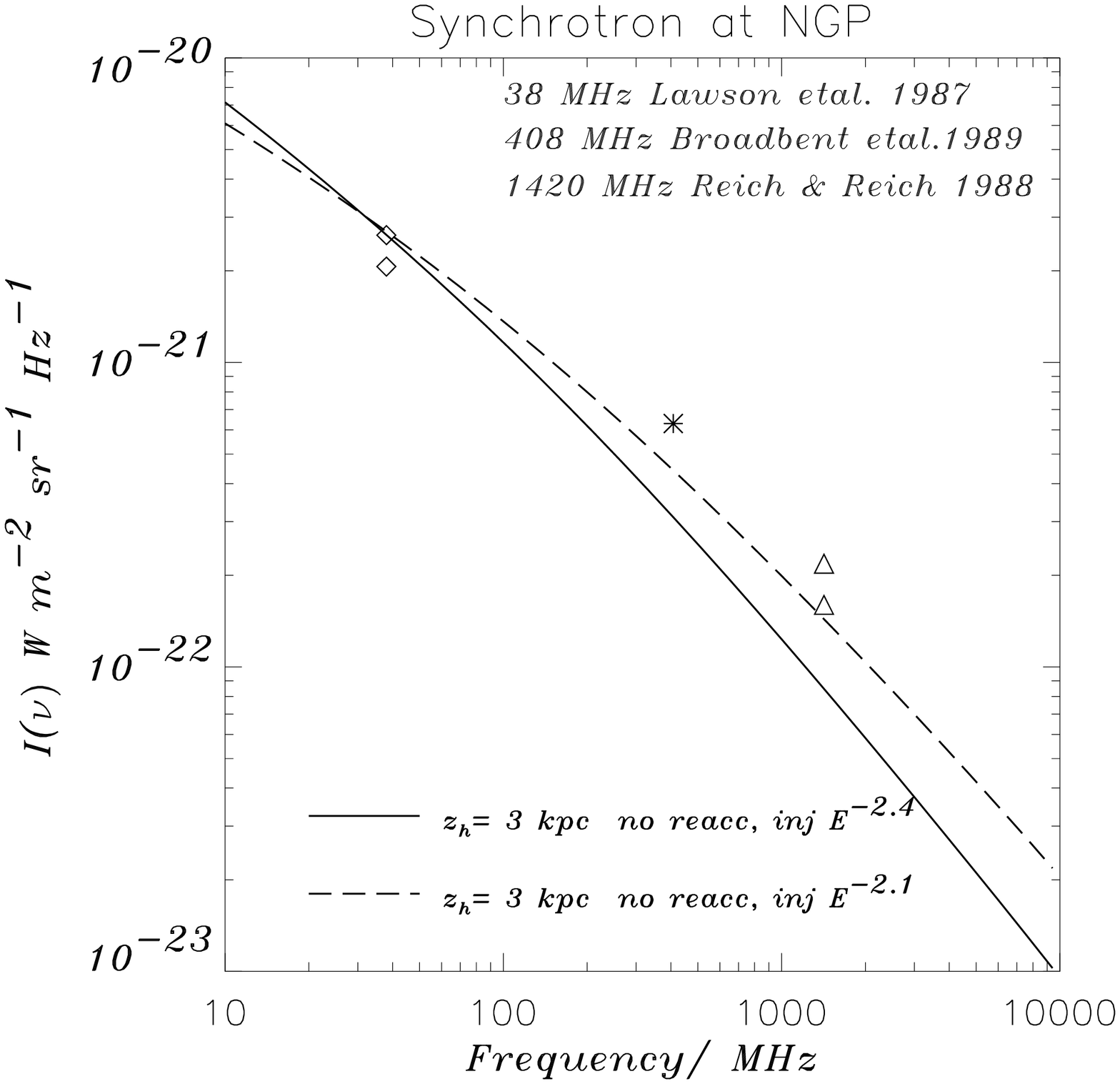,height=\figheight,width=\figwidth,clip=}}}
      \put(85,0){\makebox(80,65)[l]%
{\psfig{file=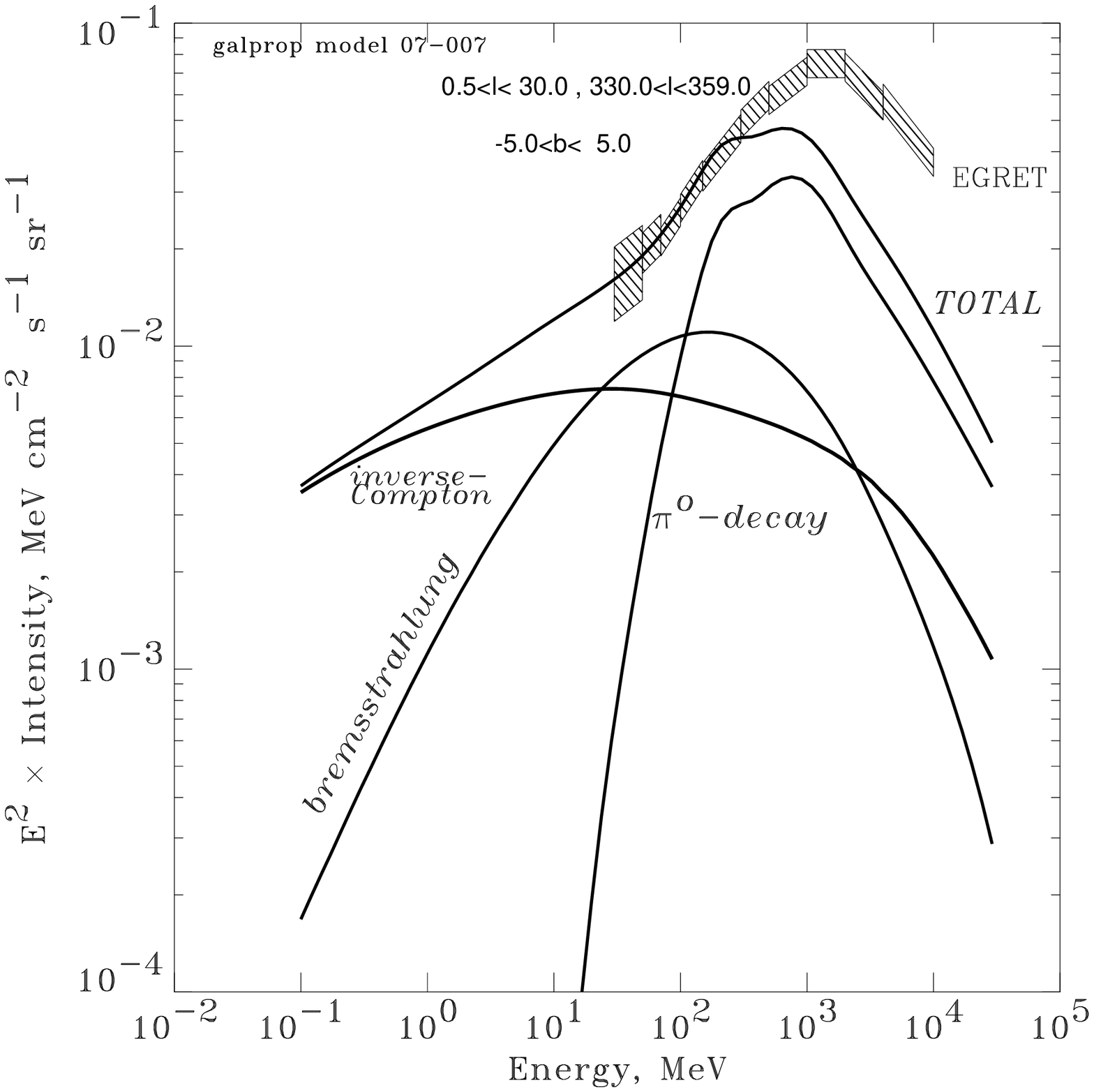,height=\figheight,width=\figwidth,clip=}}}
   \end{picture}
   \begin{picture}(170,13)(0,-3)
      \put(0,0){ \begin{minipage}{82mm} 
\it Fig.~5: The synchrotron spectrum at the NGP, compared to predictions for electron injection indices --2.1 (dashed line) and --2.4 (solid line).
        \end{minipage} }
      \put(88,0){ \begin{minipage}{82mm} 
\it Fig.~6: The $\gamma$-ray spectrum for the inner Galaxy, 
$330\degr<l<30\degr, |b|<5\degr$. EGRET data: Strong and Mattox (1996).
        \end{minipage} }
   \end{picture}
\end{figure}

\def\figheight{60mm}
\def\figwidth{60mm}
\begin{figure}[thb]
   \begin{picture}(170,62)(1,0)
      \put(0,0){\makebox(60,60)[l]%
{\psfig{file=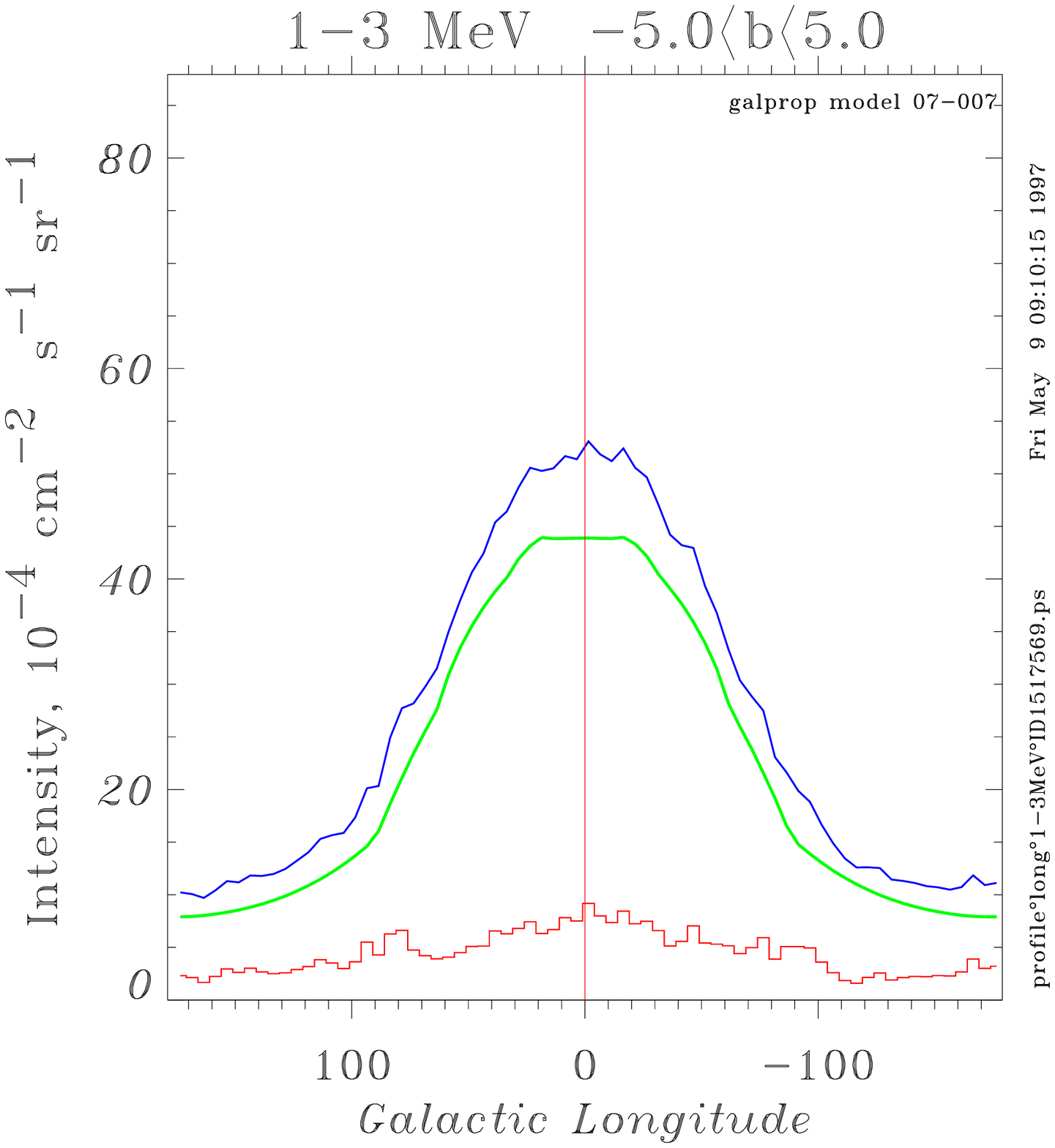,height=\figheight,width=\figwidth,clip=}}}
      \put(56,0){\makebox(60,60)[l]%
{\psfig{file=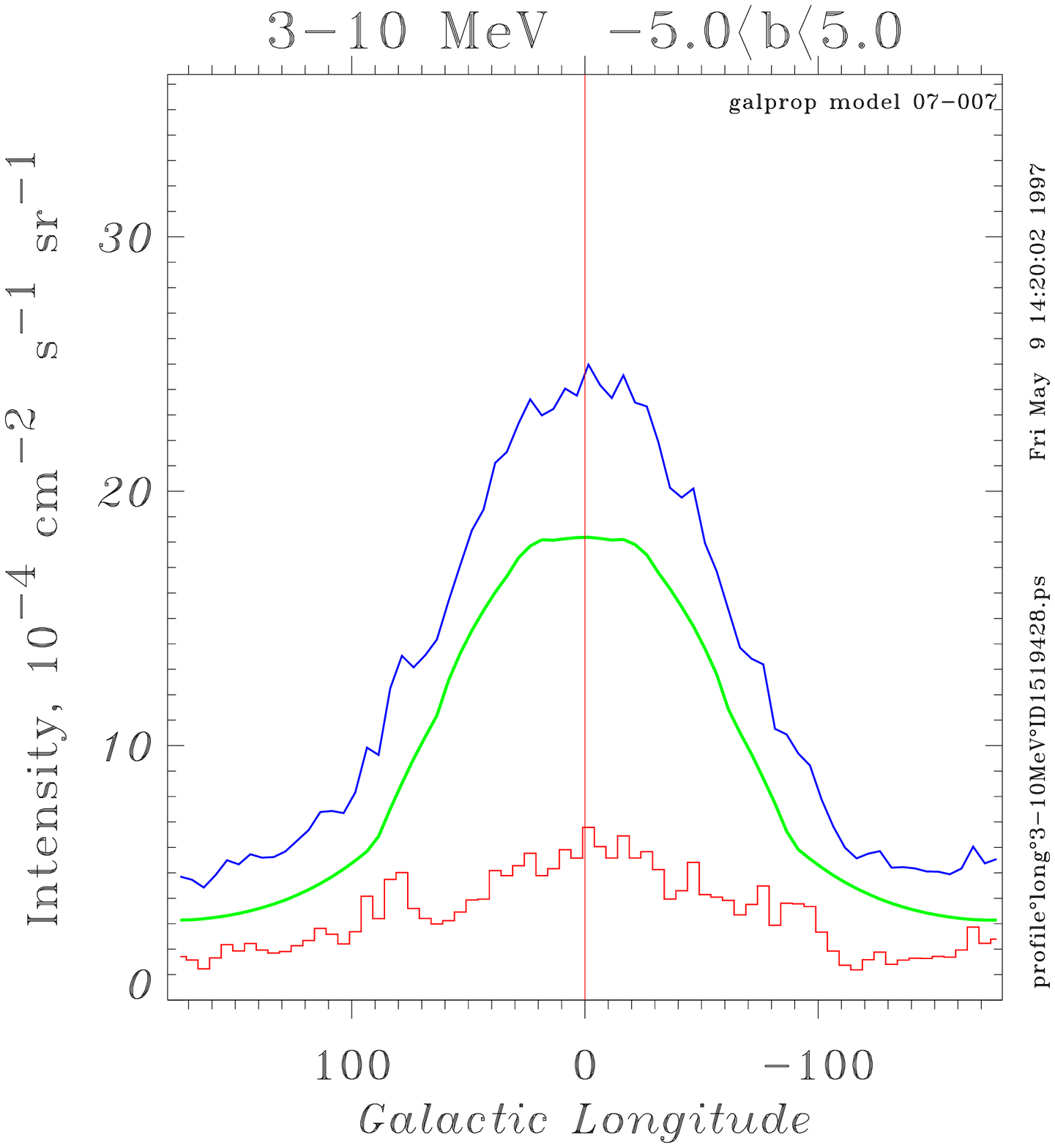,height=\figheight,width=\figwidth,clip=}}}
      \put(112,0){\makebox(60,60)[l]%
{\psfig{file=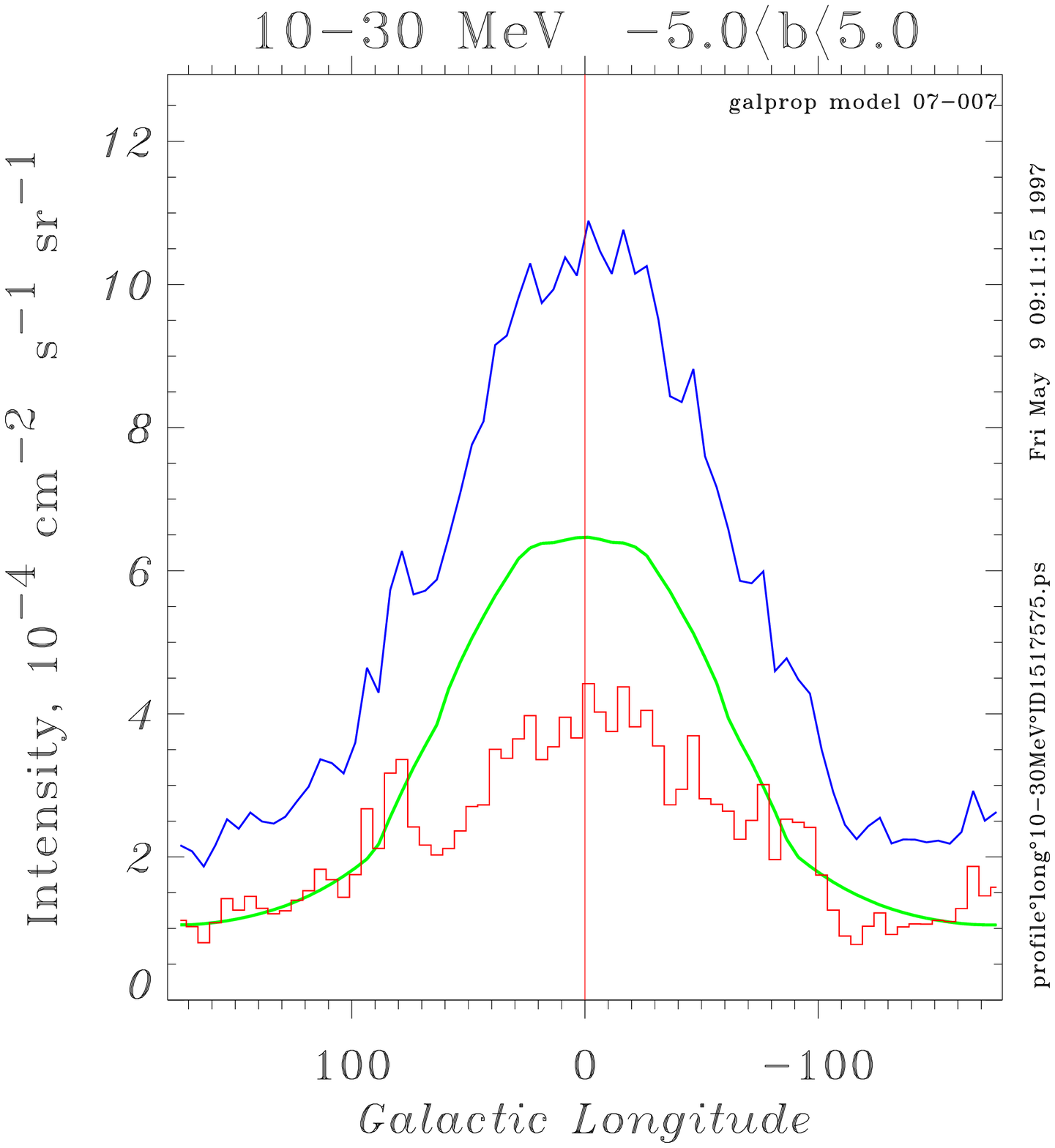,height=\figheight,width=\figwidth,clip=}}}
   \end{picture}
{\it Fig.~7: The model $\gamma$-ray longitude profile ($|b|<5\degr$)
for bremsstrahlung (histogram), inverse Compton (middle curve) and
total  (upper curve), for energies 1--3, 3--10 and 10--30 MeV.}
\vskip -4mm
\end{figure}

{\it The synchrotron spectrum}
(Fig~5) at high Galactic latitudes is important since its shape
constrains the shape of the 1--10 GeV electron spectrum.  An injection
index --2.1  (without reacceleration) is the steepest which is allowed
by the radio data over the range 38 to 1420 MHz. As illustrated, an
index --2.4  as often used (e.g.\ Strong et al.\ 1996) gives a
synchrotron spectrum which is too steep.

{\it The modelled gamma-ray spectrum}
for the inner Galaxy, illustrated here for the case of no
reacceleration (Fig~6), fits well the COMPTEL and EGRET data up to 500
MeV beyond which there is the well-known excess not accounted for by
$\pi^0$-decay with the standard nucleon spectrum.  The COMPTEL
spectrum from a recent analysis by Strong et al.\ (1997)  is both
lower and flatter than previously published. It fits well to the
flatter electron injection spectrum required by synchrotron data (see above).  
Inverse Compton is dominant below 10 MeV, bremsstrahlung
becomes important for 3--200 MeV.  The lower bremsstrahlung combined
with  $\pi^0$-decay leads to a good fit to the steep rise in this
range, in contrast to previous attempts to model the spectrum with a
steeper bremsstrahlung spectrum (Strong et al.\ 1996).  Below 10 MeV
inverse Compton becomes increasingly dominant, and bremsstrahlung is
negligible below 1 MeV.
The longitude profile (Fig~7) at low latitudes  from this model is
compared with that from COMPTEL elsewhere (Strong et al.\ 1997). The
large energy dependence of the relative importance of inverse Compton
and bremsstrahlung is evident.

More details can be found on {\it
http://www.gamma.mpe--garching.mpg.de/$\sim$aws/aws.html}

\section{REFERENCES} \footnotesize 
\setlength{\parindent}{-5mm}
\begin{list}{}{\topsep 0pt \partopsep 0pt \itemsep 0pt \leftmargin 5mm
\parsep 0pt \itemindent -5mm}
\vspace{-15pt}
   \begin{picture}(170,0)(0,0)
      \put(0,0){ \parbox[t]{75mm}{ 
\item Barwick S.W.\ et al., ApJL, 482, L191 (1997)
\item Dermer C.D., A\&A, 157, 223 (1986) 
\item Engelmann J.J.\ et al., A\&A, 233, 96 (1990) 
\item Heinbach U., and Simon M., ApJ, 441, 209 (1995)
\item Mori M., ApJ, 478, 225 (1997) 
\item Moskalenko I.V., and Strong A.W., in preparation (1997) 
\item Protheroe R.J., ApJ, 254, 391 (1982)
\item Seo E.S., and Ptuskin V.S., ApJ, 431, 705 (1994) 
\item Seo E.S.\ et al., ApJ, 378, 763 (1991) 
}}
      \put(85,0){ \parbox[t]{75mm}{
\item Strong A.W., and Mattox J.R., A\&A, 308, L21 (1996)
\item Strong A.W., and Moskalenko I.V., Proc.\ 4th Compton Symp., AIP (1997) 
\item Strong A.W., and Youssefi G., Proc.\ 24th ICRC 3, 48 (1995)
\item Strong A.W.\ et al., Proc.\ 3rd Compton Symp., A\&A Supp 120,
 C381 (1996)
\item Strong A.W.\ et al., Proc.\ 4th Compton Symp., AIP (1997)
\item Webber W.R.\ et al., ApJ, 457, 435 (1996)
}}
   \end{picture}

\end{list}

\end{document}